# Rheological behavior of molybdenum disulfide (MoS$_2$) inks under electric fields: influence of concentration and voltage

*Pedro C. Rijo[1,2] and Francisco J. Galindo-Rosales[2,3]\**

[1]Transport Phenomena Research Center (CEFT), Mechanical Engineering Department, Faculty of Engineering, University of Porto, Rua Dr. Roberto Frias, 4200-465 Porto, Portugal.

[2]AliCE – Associate Laboratory in Chemical Engineering, Faculty of Engineering, University of Porto, Rua Dr. Roberto Frias, 4200-465 Porto, Portugal.

[3]Transport Phenomena Research Center (CEFT), Chemical Engineering Department, Faculty of Engineering, University of Porto, Rua Dr. Roberto Frias, 4200-465 Porto, Portugal.

\*Corresponding Author: galindo@fe.up.pt

**Abstract**

This work provides a complete rheological characterization of molybdenum disulfide (MoS$_2$) inks in the presence of electric fields. Several concentrations of MoS$_2$ are studied and dispersed in a viscoelastic fluid. The lubrication effects are present in the ink when the MoS$_2$ concentration is higher than 0.10% w/w. The dielectric properties show the impossibility of a positive electrorheological effect for all MoS$_2$-inks studied. The formation of vortices and electromigration of MoS$_2$ particles occur under the influence of an external electric field. These two phenomena affect the rheological behavior of MoS$_2$-inks under shear flow condition. Relatively to the extensional rheology experiments, the particle migration and the vortex formation promote anisotropy on the rheological properties of the inks which affects the relaxation time, the formation of beads-on-a-string and the uniaxial elongational flow condition is no longer valid. When the electric field strength is 1.5 kV/mm, the formation of Taylor's cone is observed and independent of MoS$_2$ concentration.

# 1. Introduction

Molybdenum disulfide ($MoS_2$) is a semiconductor material belonging to the transition metal dichalgenides family [1, 2]. This material has aroused interest in the scientific commnunity due to its optoelectronic properties due to the arrangement of the molybdenum (Mo) and sulfur (S) atoms in a two-dimensional plane similar to a honeycomb structure. $MoS_2$ has been used to produce sensors, thin film transistors, photodetectors, and supercapacitors [1, 2].

To produce these microelectronic devices several techniques, such as lithography, screen gravure, inkjet printing, electrophoretic deposition (EPD) and electrohydrodynamic (EHD) jet printing have been developed[1-4]. These techniques differ considerably in technical characteristics and capabilities. If ultra-high-resolution is a requirement for the final product, then the EHD jet printing is the most suitable technique [5]. This technique is a non-vacuum, direct-writing, non-contact, and low cost process that can be opperated either in drop-on-demand or in continuous jet [3]. Independently of the printing process, the viscosity is one of the parameters, in addition to electrical conductivity and surface tension, can dictate which printing process is more suitable for obtaining high-resolution products [10].

Using 2D nanoparticles, Marra *et al.*[6] found that the shear-thinning behavior of graphene-inks is more notorious than that observed for the dispersant medium at high shear rates.. Further, the viscosity of graphene-ink is lower than that of the graphene-free ink in the shear-thinning zone due to the lubrication effects of the graphene particles. According to Kamal *et al.*[7], the lubricant effect or hydrodynamic slip effect occurs when a rigid graphene nanoplatelet does not rotate in the liquid when the shear flow is imposed and it gets aligned with the flow direction reducing the resistance of the fluid to flow. Further, Ivanova *et al.*[8] found that slip can be controlled by changing the wettability of the surface. So, non-wetted or partially wetted surfaces of graphene nanoplatelets (GNP) promote slip, while wetted surfaces with strong fluid-surface interactions prevent slip.

In $MoS_2$ suspension, Wan *et al.*[9] found the hydrodynamic slip effects when these nanoparticles were dispersed in Newtonian fluids. Moreover, Rijo *et al.* [10] characterized the rheological properties of three different 2D particles, i.e. $MoS_2$, GNP and hBN (hexagonal boron nitride), dispersed in a Newtonian and a viscoelastic fluids. The authors found that their rheological behaviors did not change when the particle concentration is very small in steady-shear flow condition; however, regarding the uniaxial elongation flow, the results showed that pinch-off phenomenon is present even when particles are dispersed in Newtonian fluid and the formation of beads-on-a-string structures is affected by the presence of 2D particles in the viscoelastic suspending fluid. Further, the extensional relaxation time of the ink reduces when 2D particles are present.

When the electric field is applied to the fluid, Lee *et al.* [11], Mrlik *et al.* [12] and Yin *et al.* [13] studied the electrorheological effect of 2D particles dispersed in a Newtonian fluid. They found that the presence of 2D particles improves the dielectric properties since the polarization effect is the main mechanism responsible for the appearance of ER effect. Moreover, they observed that the temperature and the modification of the particle surface affect the ER effect.

Recently, Rijo and Galindo-Rosales [14] studied the rheological properties of 2D inks when the electric field was parallel to or perpendicular to the flow field. They found that, at low concentration (0.2 mg mL$^{-1}$), the 2D nanoparticles did not alter the flow behavior of the carrier fluid under shear flow conditions, i.e. the suspension remained Newtonian when dispersed in a Newtonian fluid, even when the electric field was applied, and the shear behavior remained shear-thinning when dispersed in a viscoelastic fluid. Relatively to extensional rheological characterization, the authors found that the electric field does not affect the Newtonian fluid condition when the particles are dispersed in Newtonian fluid. However, the filament thinning process is affected by the kind of 2D nanoparticle and the intensity of the electric field due to electrophoresis and vortex formation appear, which affects the relaxation time of the inks.

The previous work [14] studied how the electric conductivity of 2D particles, at low particle concentration, influence the the rheological properties when an electric field is applied. The present work has the goal of further studying how the 2D nanomaterial concentration affects the rheological properties on the presence of electric field. The rheological characterization will be done when the electric field is parallel to or perpendicular to the flow field.

## 2. Materials and Methods

### 2.1. Materials

The dispersant/carrier fluid was a 2.5% w/v polymer solution of ethyl cellulose (48% ethoxyl basis, Acros Organics) dissolved in toluene (99% purity, Carlos Erba Reagents). $MoS_2$ nanoparticles (Sigma-Aldrich) were dispersed at different concentrations (**Table 1**), following the same preparation protocol and the formulation as those used in [10].

**Table 1.** Density (ρ) and surface tension (σ) of each suspension.

| $\phi_{MoS_2}$ [% w/w] | $\rho$ [kg/m$^3$] | $\sigma$ [mN/m] |
|---|---|---|
| 0 | 871.45 | 27.80 ± 0.01 |
| 0.023 | 874.54 | 27.96 ± 0.03 |
| 0.10 | 874.97 | 28.15 ± 0.04 |
| 0.25 | 874.13 | 28.55 ± 0.02 |
| 0.50 | 879.79 | 28.56 ± 0.03 |
| 0.75 | 878.83 | 28.34 ± 0.01 |
| 1.5 | 894.90 | 25.57 ± 0.02 |
| 3.0 | 900.46 | 25.35 ± 0.01 |

## 2.2. Dielectric Properties

The dielectric properties were measured using a Keysight E4980AL LCR meter and the measurements were done at 20 °C and 1 V in an alternating current (AC) frequency window between 50 Hz and 500 kHz. Two parallel plates of stainless steel (50 mm in diameter) were used, filling the empty gap with fluid when the plates were 0.5 mm away from each other.

## 2.3. Rheological Measurements

### 2.3.1. Rheology and electrorheology under shear flow

The shear viscosity was measured using an electrorheological device (ERD) coupled to a controlled shear-stress rheometer (Anton Paar MCR301) (**Figure 1**). The plate-plate geometry of 50 mm diameter was used with a gap of 0.1 mm connected to a high voltage power supply (FUG HCL 14-12500). Five independent runs were carried out for each fluid to ensure good repeatability of results at a temperature of 20 °C.

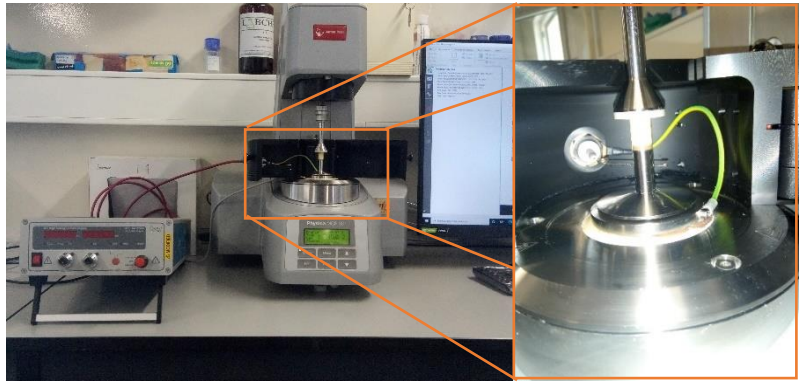

**Figure 1.** Experimental setup used to measure steady-shear viscosity, and the electrorheological cell used to apply electric field to the fluids. Reproduced with permission of [14].

*2.3.2. Rheology and electrorheology under extensional flow*

The extensional rheological characterization was performed using a capillary breakup extensional rheometer (CaBER) coupled with a high-speed camera (Photron FASTCAM mini UX100) and a light source (Leica EL6000). To see the filament thinning process in detail, we used a set of optical lenses (Optem Zoom 70XL) with a variable magnification from 1X to 5.5X. **Figure 2b** and **c** show the plates used to study the filament thinning process with and without application of and electric field. The experimental protocol used in this work is the same as used in [14].

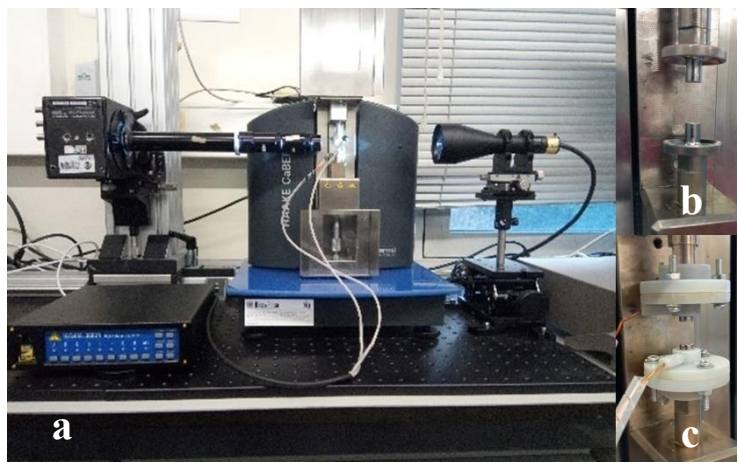

**Figure 2. (a)** Experimental setup used for extensional tests. **(b)** Standard 4 mm plate geometry used during the experiments without the application of an external electric field. **(c)** Electrified 4 mm plate geometry used during the experiments when an electric field is applied. Reproduced with permission of [14].

## 3. Results and Discussion

### 3.1. Dielectric Properties of the MoS$_2$-inks

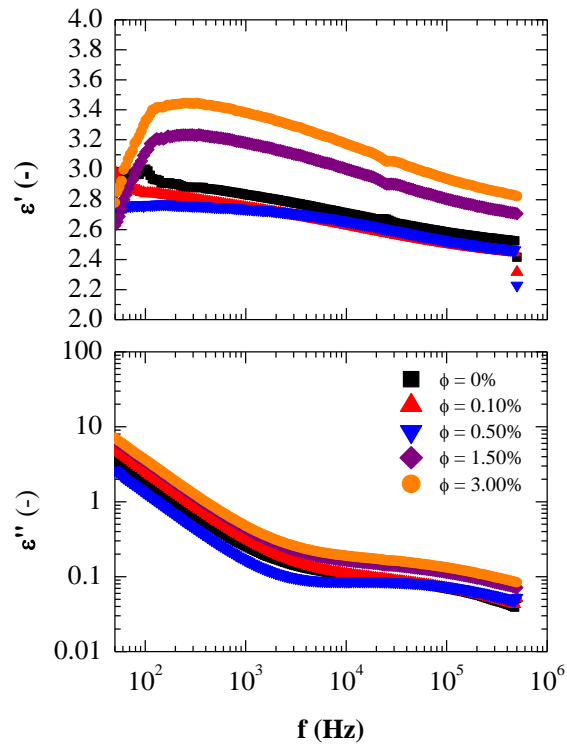

**Figure 3.** Dielectric constant (ε') and dielectric loss factor for five different MoS$_2$-inks

**Figure 3** shows the dielectric constant and dielectric loss factor curves for five different MoS$_2$ concentrations: 0%, 0.10%, 0.50%, 1.50% and 3.00%. The curves for the concentrations 0.023%, 0.25% and 0.75% have been removed from **Figure 3** and placed in **Figure S1** (Support Information) in order to simplify the analysis of the results and the visualization of the curves obtained for the dielectric constant and dielectric loss. Regarding the polymeric solution of 2.5% w/v ethyl cellulose dissolved in toluene, the curves present in both figures follow the same trend verified by Nojiri and Okamoto [15] for the poly(vinyl acetate)-toluene solution. The authors verified that the temperature of solution, the concentration and the molecular weight of polymer influence the dielectric properties. Unlike in suspensions, Maxwell-Wagner polarization is not the mechanism responsible for the behavior of the curves obtained in **Figure 3** for the carrier fluid, but rather the segmental relaxation (α-relaxation) process [15]. This process controls the diffusion, viscosity and rotation of the monomers and it is usually ascribed to micro-Brownian motion of the chain segments.

When MoS$_2$ particles are added to the suspension and regardless of the particle concentration used, the relaxation frequency is found to be below 100 Hz, and thus outside the desirable frequency range (100 Hz to 100 kHz). This shows that the polarization rate is so low that it does not allow the formation of particle structures that promote a positive electrorheological effect, i.e., an increment of viscosity in the presence of electric field. Furthermore, the difference in dielectric constants (Δε') between 100 Hz and $10^5$ Hz calculated for all concentrations (**Table 2**), and Δε' is less than 0.421.

**Table 2.** Evolution of dielectric constant drop (Δε'= ε'($10^2$Hz) - ε'($10^5$Hz)) with $MoS_2$ nanoparticle concentration at 20 °C.

| $\phi_{MoS_2}$ [% w/w] | Δε' [-] | $\phi_{MoS_2}$ [% w/w] | Δε' [-] |
|---|---|---|---|
| **0** | 0.421 | **0.50** | 0.239 |
| **0.023** | 0.349 | **0.75** | 0.354 |
| **0.10** | 0.338 | **1.50** | 0.318 |
| **0.25** | 0.097 | **3.0** | 0.393 |

These results contradict the formation of columnar structure of $MoS_2$ particles dispersed in silicone oil when an electric field is applied, as reported by Lee *et al.* [11]. The authors used silicone oil as carrier fluid with a zero-shear viscosity of 100 cSt, which is equivalent to 96 mPa·s, and a dielectric constant of 2. While the solution of 2.5% w/v of ethyl cellulose dissolved in toluene shows a zero-shear viscosity of 18.64 ± 0.42 mPa·s and a dielectric constant above 2.25 and independently of the $MoS_2$ concentration for the frequency range between 100 Hz and $10^5$ Hz. It is well known that nanocomposites exhibit different rheological behavior depending on the quality of the dispersion [16]. Electrorheology is not an exception, having in both systems (silicone and toluene) the same degree of dispersion may provide a more uniform and predictable response. Thus, we could speculate that the dispersion quality may be responsible for such an apparent contradiction between the two electrorheological effects reported in these two research studies. However, in order to be conclusive, a systematic research study similar to that developed by Galindo-Rosales *et al.* [17] would be required, which is out the scope of the current study.

### 3.2. Shear Rheological Characterization

The effect of concentration of $MoS_2$ particles for several electric field strengths is shown in **¡Error! No se encuentra el origen de la referencia.** for five different $MoS_2$ concentrations: 0%, 0.10%, 0.50%, 1.50% and 3.00%. The shear stress curves for the concentrations of 0.023%, 0.25% and 0.75% have been removed from **Figure 4** and placed in **Figure S2** (Support Information) to simplify the analysis of the results and the visualization of the shear stress curves obtained for different electric field strengths. In absence of electric field (**Figure 4a**), it is possible to observe that the particle concentration above 0.10% w/w affects significantly the flow curves for shear rates values above 100 s$^{-1}$. The shear stress is lower than the one obtained for the carrrier fluid (ϕ = 0% w/w). The possible cause for this is the lubrication effect of MoS$_2$ particles and it was also observed by Wan *et al.* [18] According to Panitz *et al.* [19], the lubrification effect of MoS$_2$ is associated to weak van der Waals' bonding between adjacent basal planes of the hexagonal structure that permits shear displacement at low shear stress.

As reported by Rijo and Galindo-Rosales [14] for a suspension of 0.023% w/w MoS$_2$, when an electric field is applied to the fluid, two physical mechanisms appear and affect the flow curves: electrophoresis and vortex formation. According to Barrero *et al.* [20], the formation of vortices inside of the fluid can be derived by the tangential electrical stresses acting on the liquid-gas interface; moreover, the intensity of the vortices depends on the electrical conductivity and viscosity of the fluid, being the vortices more intense when the fluid exhibits both low electrical conductivity and low viscosity. This phenomenon can also be observed in Taylor's cones in the presence of electric field at null flow rate [21].

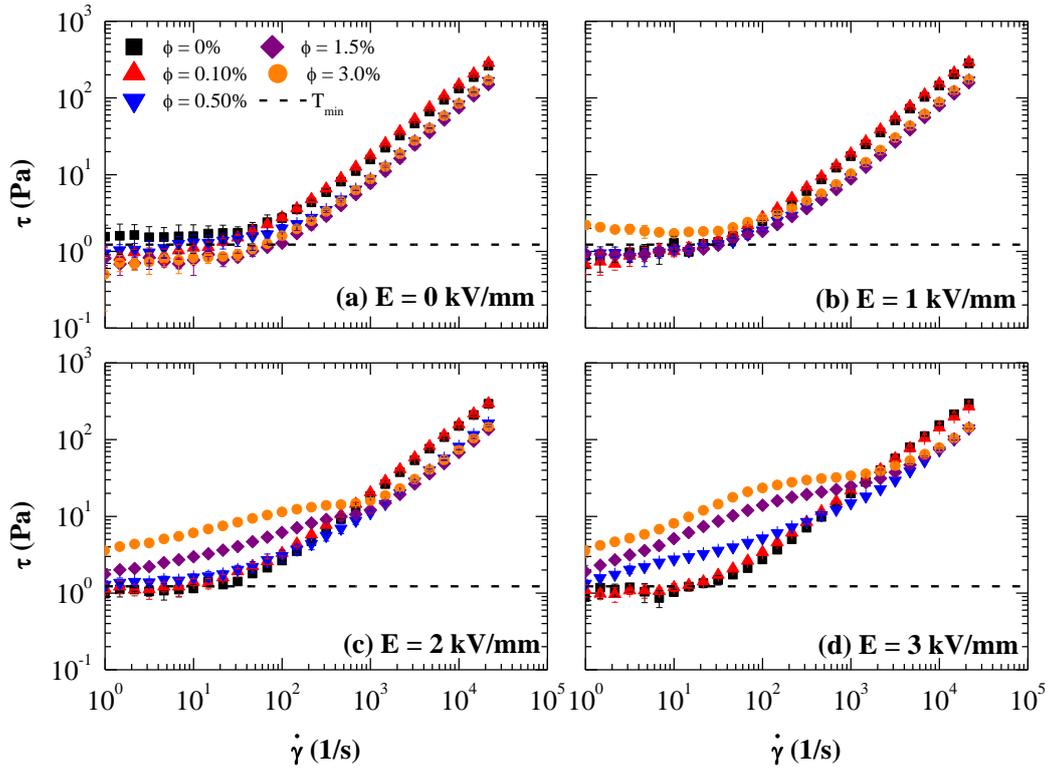

**Figure 4.** Influence of MoS$_2$ concentration (ϕ = 0%, 0.10%, 0.50%, 1.5% and 3.0% w/w) on the shear stress curves for: **(a)** E = 0 kV/mm, **(b)** E = 1 kV/mm, **(c)** E = 2 kV/mm, and **(d)** E = 3 kV/mm.

The electrophoresis mechanism also occurs for higher concentrations and it is responsible for the phase separation between the MoS$_2$ particles and the carrier fluid, as shown in the sequence of images reported in **Figure 5** and in **Video S1** in Support Information.

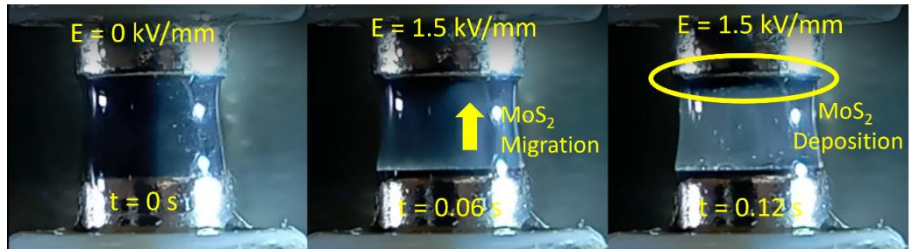

**Figure 5.** MoS$_2$ migration of 0.50% w/w MoS$_2$ dispersed in 2.5% w/v EC+Tol when an electric field of 1.5 kV/mm was applied. (Gap between two plates is 2 mm)

Considering these two mechanisms, **Figure 4 b-d** show an increment of shear stress at low shear rates when an electric field is applied to the fluid; moreover, the increment of the shear stress is more notorious when $\phi \geq 0.50\%$ and $E = 3$ kV/mm. This increment may be due to the randomly deposition of MoS$_2$ particles on the surface of the positive electrode, which establishes contact points with counter electrode. The establishment of these contact points is due to the small distance between the plates used during the electrorheological characterization. These contact points are destroyed as the shear rate increases and promotes an efficient deposition of MoS$_2$ particles on the positive electrode aided by the fluid vortices. As the distance between the plates is so small, there is a false positive electrorheological effect. When the gap increases to 0.250 mm **(Figure 6a)**, this distance prevents the establishment of contact points between two electrodes during the particle migration and the flow curve at low shear rates follows the behavior observed when $\phi < 0.50\%$ and $E = 3$ kV/mm. **Figure 4 b-d** shows that all contact points are destroyed when the shear rate overcomes 1000 s$^{-1}$ and the particles are deposited in compact form on the electrode. When this happens, the shear stress for suspensions with $\phi \geq 0.50\%$ are lower than the shear stress measured for the carrier fluid.

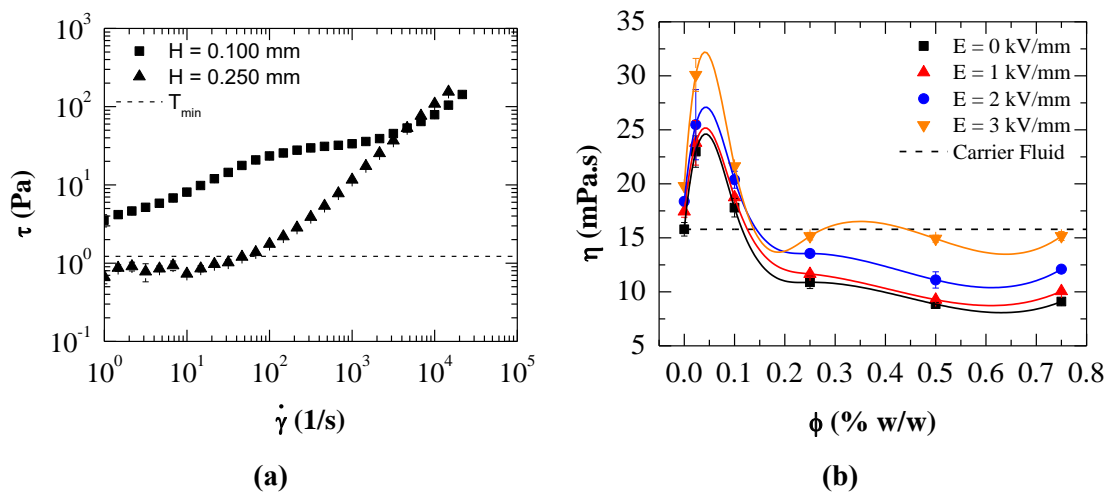

**Figure 6. (a)** Dependency of electrorheological effect on the distance between the plates for 3.0% w/w MoS$_2$ and E = 3 kV/mm. **(b)** Influence of MoS$_2$ concentration on the steady-shear viscosity for a constant shear rate ($\dot{\gamma} = 1000\ s^{-1}$). Solid lines are guides for the eye.

When the particle concentration is lower or equal to 0.10% w/w the suspensions have a shear stress higher than or close to the shear stress measured for the carrier fluid when the shear rates are higher than 100 s$^{-1}$ and independently of the electric field strength (**Figure 4**). This means that exists a minimum/critical concentration of MoS$_2$ particles from which the lubrication effect become relevant. **Figure 6b** shows that the critical concentration is $0.13 \pm 0.02\%$, above which the lubrication effect starts reducing the viscosity of the fluid below the one of the carrier fluid. **Figure 6b** also allows to infer the pure effect of the recirculations on the values of the viscosity, which increases up to a 25% when no particles are dispersed into the system.

### 3.3. Extensional Rheological Characterization

For the sake of simplicity, this section will only reproduce the extensional viscosity curves and the evolution of the liquid filament radius over time for MoS$_2$-inks with conctretations of 0%, 0.10%, 0.50%, 1.5% and 3.0% respectively. For the other concentrations, the results obtained are reproduced in **Figures S3** and **S4** in the Support Information.

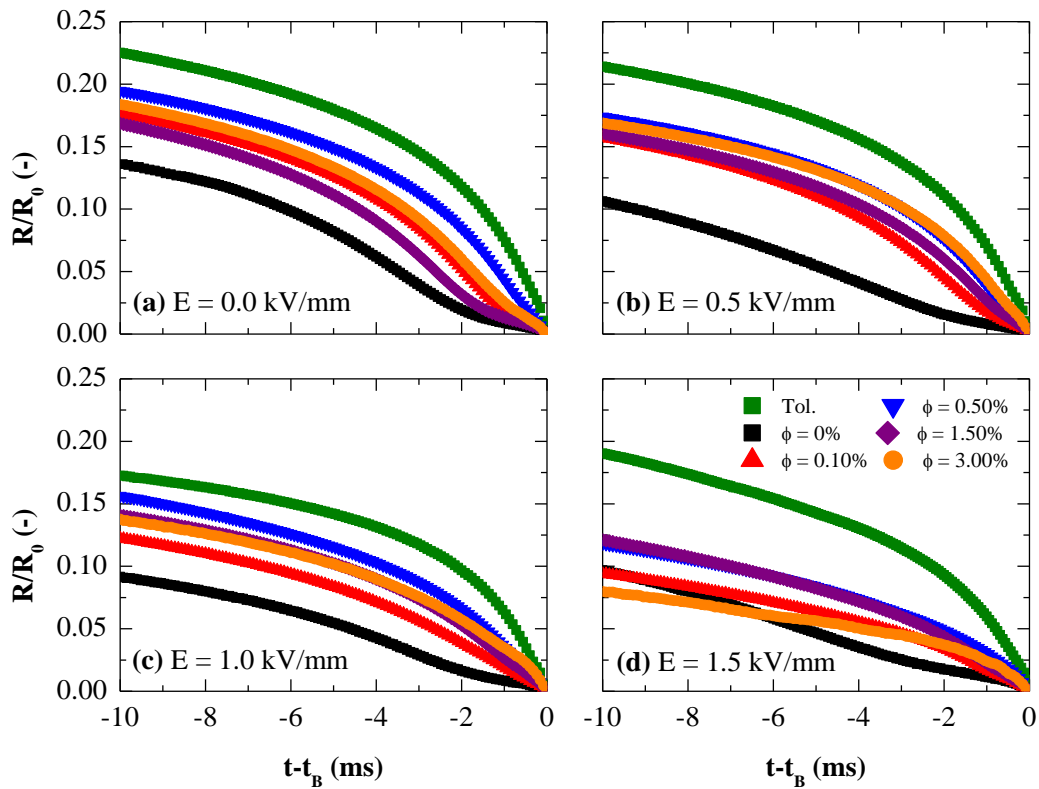

**Figure 7.** The normalized radius decay profiles for five different MoS$_2$-inks ($\phi$ = 0%, 0.1%, 0.5%, 1.5% and 3.0% w/w) and toluene for different electric field strengths: **(a)** 0 kV/mm, **(b)** 0.5 kV/mm, **(c)** 1.0 kV/mm and **(d)** 1.5 kV/mm.

**Figure 7** shows that the addition of polymer and MoS$_2$ nanoparticles increase the resistance of the fluid to thin in comparison with the pure toluene case. In the case of toluene, the Ohnesorge number is lower than 0.2077 [22] which evidences that the capillary thinning process is influenced by the inertia present in the system. When polymer or polymer + particles are added to toluene, the Ohnesorge number is higher than 0.2077, which shows that the thinning process is controlled by the viscosity and surface tension of the fluid. Independently of the electric field strength, the polymer solution provides greater fluid stretching resistance when compared to fluid containing MoS$_2$ particles dispersed in 2.5% w/v ethyl cellulose + toluene (**Figure 7**).

**Table 3** – Relaxation times of MoS$_2$ particles dispersed in 2.5% w/v ethyl cellulose + toluene under the application of an external electric field aligned with the direction of the flow.

| $\phi_{MoS_2}$ [% w/w] | 0 kV/mm $\lambda$ [ms] | 0.5 kV/mm $\lambda$ [ms] | 1.0 kV/mm $\lambda$ [ms] | 1.5 kV/mm $\lambda$ [ms] |
|---|---|---|---|---|
| 0 | 0.451 ± 0.041 | 0.475 ± 0.038 | 0.474 ± 0.061 | 0.444 ± 0.022 |
| 0.023 | 0.174 ± 0.003 | 0.208 ± 0.003 | -- | -- |
| 0.10 | 0.274 ± 0.009 | -- | 0.257 ± 0.014 | -- |
| 0.25 | 0.240 ± 0.008 | 0.127 ± 0.004 | -- | -- |
| 0.50 | 0.172 ± 0.003 | 0.144 ± 0.002 | 0.245 ± 0.008 | |
| 0.75 | 0.185 ± 0.006 | -- | -- | -- |
| 1.5 | 0.403 ± 0.046 | 0.283 ± 0.009 | 0.355 ± 0.020 | 0.344 ± 0.026 |
| 3.0 | 0.308 ± 0.022 | -- | 0.528 ± 0.010 | -- |

In the absence of electric field, the relaxation time of MoS$_2$-inks decreases compared to carrier fluid (see **Table 3**). This behavior is contrary to what was observed by Ock *et al.* [23], who studied the capillary thinning process with PMMA particles added to a polyethylene oxide solution. They found that spherical PMMA particles increased the thinning rate and extensional relaxation time compared to the pure polymer solution. However, MoS$_2$ particles are two-dimensional and can agglomerate within the suspension. At the microscopic level, MoS$_2$ particles may promote local shear rates during capillary thinning, restricting the uncoiling of ethyl cellulose polymer chains. Additionally, MoS$_2$ particles' lubricating effect could limit uncoiling during particle-polymer interactions, contributing to a reduction in relaxation time. This slipping phenomenon was also observed by Kamal *et al.* [7] in a simple shear flow condition. At the macroscopic level, the presence of partiles in fluids regardless of aspect ratio promotes the same behavior observed by Mathues *et al.* during the filament stretching and breakage [24].

Considering the reasoning of Mathues *et al.*, it is possible to observe the deccelerated regime in the MoS$_2$-inks for the last 2 ms of the filament lifetime (**Figure 7**). In this regime, the filament radius decay can be described by a exponential law equation used to describe the filament radius decay of a viscoelastic fluid. The exponential law is described as [25]:

$$\frac{R_{min}}{R_0} = \left(\frac{GR_0}{2\sigma}\right)^{1/3} \exp\left(-\frac{t}{3\lambda}\right) \quad (1)$$

where $R_0$ is the initial radius of the filament, $G$, $\sigma$ and $\lambda$ are the shear modulus, surface tension and extensional relaxation time of the fluid, respectively. In this region, the elastic fluid condition is respected and the relaxation time of the suspensions was calculated using the **equation 1** and **Table 3** shows the relaxation times for each suspension. Moreover, time interval used to calculate the relaxation time is the same where an increase in the extensional viscosity can be observed for a constant extension rate or for a very slow increase in the extension rate (**Figure 8).** This increase in viscosity occurs when the elastic forces overcome the viscous ones, which corresponds to the uncoiling of the polymer chains [26].

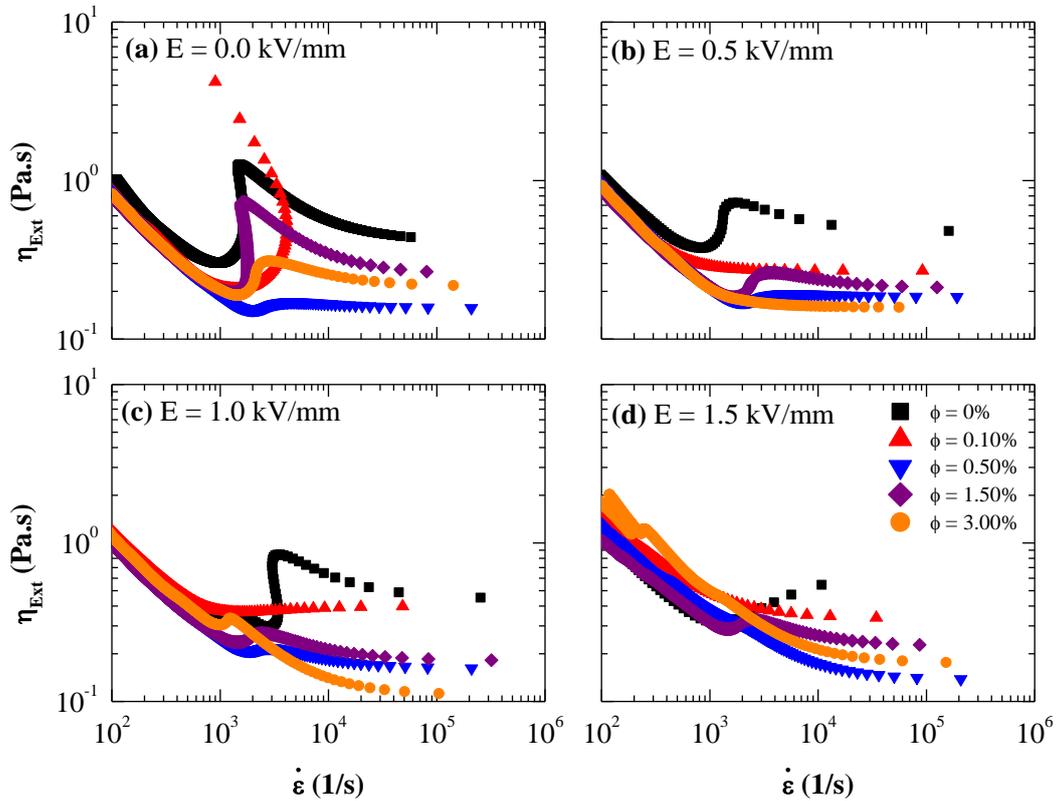

**Figure 8.** The apparent extensional viscosity ($\eta_{Ext}$) as a function of extension rate ($\dot{\varepsilon}$) for five different MoS$_2$-inks ($\phi$ = 0%, 0.1%, 0.5%, 1.5% and 3.0% w/w) for different electric field strengths: **(a)** 0 kV/mm, **(b)** 0.5 kV/mm, **(c)** 1.0 kV/mm and **(d)** 1.5 kV/mm.

The presence of an electric field promotes the appearance of two distinct phenomena, independently on the flow/field configuration: vortex formation and electromigration of MoS$_2$ particles to the positive electrode through electrophoresis. During the capillary thinning process, these two phenomena alter the rheological behavior of the fluid under test. The presence of vortices promotes the appearance of local shear rates that can influence the uncoiling of the ethyl cellulose polymer chains which in turn influences the extensional relaxation time of the fluid; moreover, the presence of these two phenomena invalidates the uniaxial elongational flow condition. Thus, in this case, the calculated λ parameter corresponds to an apparent relaxation time. When the applied electric field strength increases, the filament thinning slows down (**Figure 9a**). One might expect the presence of vortices to accelerate the thinning rate; however, this does not happen since there is a possibility that the vortices are feeding the filament prolonging its lifetime [14] (**Figure 9b**).

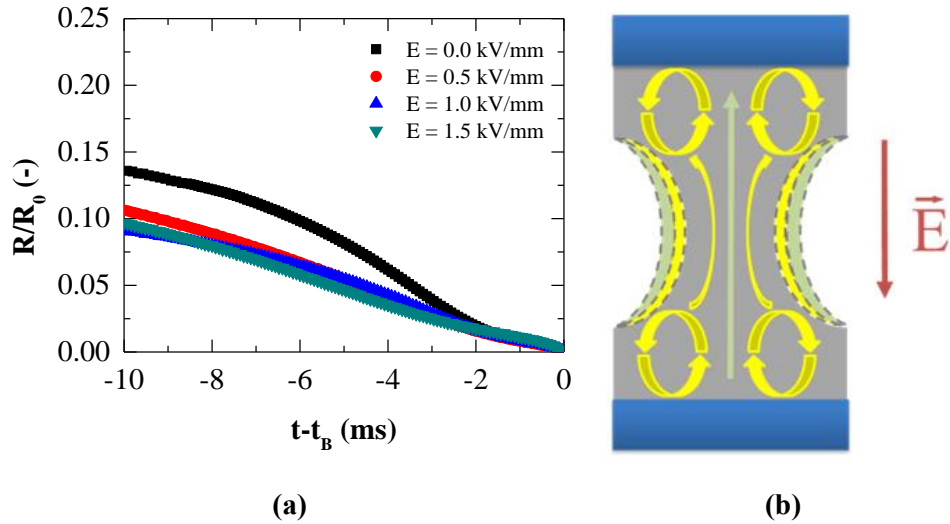

**Figure 9. (a)** Radius decay profiles for the carrier fluid of $MoS_2$-inks in the presence of electric field. **(b)** Sketch on the influence of vortex formation (yellow) and drag flow (light green) due to particle migration when the electric field is turned on. Reproduced with permission [14].

Regarding the relaxation times measured for the polymer solution, **Table 3** shows that the relaxation time is almost constant and independent of the applied electric field, which shows that the stretching of the polymer chains is not influenced by the presence of vortices caused by the application of electric field. When $MoS_2$ particles are added to the fluid, the rheological properties of the fluid change. Based on **Figure 7**, the concentration of $MoS_2$ particles influences the capillary thinning process, as well as the strength of the electric field applied to the fluid in the same direction as the flow field. Furthermore, the minimum filament radius decay curves of the inks follow the trend observed for pure toluene. This possibly happens because when an electric field is applied, the $MoS_2$ particles tend to migrate to the positive electrode, dragging with them the polymer presents in the fluid. It is expected that the higher the particle concentration, the greater the amount of polymer that is dragged to the upper plate, which can significantly change the rheological properties of the fluid.

Without electric field application, a drastic reduction in fluid relaxation time was observed due to the presence of local shear rates and possibly slipping phenomena between the particles and the fluid during the capillary thinning process. With the application of the electric field, the local concentration of available polymer in the thinning zone varies depending on the amount of ethyl cellulose entrained during the migration of $MoS_2$ particles. This new concentration value will influence the extensional relaxation time of the fluid. Based on **Table 3,** there is a certain anisotropy of the experimentally calculated values depending on the concentration of $MoS_2$ and the strength of the electric field applied to the fluid. This anisotropy of the suspensions is also reflected in the extensional viscosity curves present in **Figure 8** where the increment of the extensional viscosity for a constant or quasi-constant extension rate can be suppressed depending on the electric field strength.

This anisotropy can be explained due to the polydispersity of the size of the $MoS_2$ aggregates dispersed in the fluid. This polydispersity of the particles does not guarantee that the mesh formed during the migration of $MoS_2$ particles has the same pore size. This causes that for certain test conditions, the pore size varies randomly, thus influencing the local concentration of polymer in the thinning zone which directly influences the relaxation time of the fluid. It is important to note that the criterion used to determine the fluid relaxation time consists in fitting the evolution of the minimum filament radius through **equation 1** to the time interval where an increase in the extensional viscosity of the fluid occurs.

**Figure 10** shows the last photography for each $MoS_2$-ink immediately before the filament breakage. When the particle concentration increases in the absence of electric field, the formation of beads-on-a-string (BOAS) with a perfect viscoelastic droplet in the middle of the filament is replaced by an elongated ellipsoid droplet with a weak elasticity effect. This weak effect is clear on the relaxation times present in **Table 3**. When the electric field is switched on, the formation of BOAS tends to disappear when the $MoS_2$ concentration is null; moreover, in the presence of $MoS_2$ particles, the formation of an elongated ellipsoid droplet with a weak elasticity effect occurs and it is very similar to a pinch-off breakage of Newtonian fluids. ¡Error! No se encuentra el origen de la referencia. to ¡Error! No se encuentra el origen de la referencia. in supporting information shows the photographic record of the filament thinning process for each $MoS_2$-ink under the effect of an electric field.

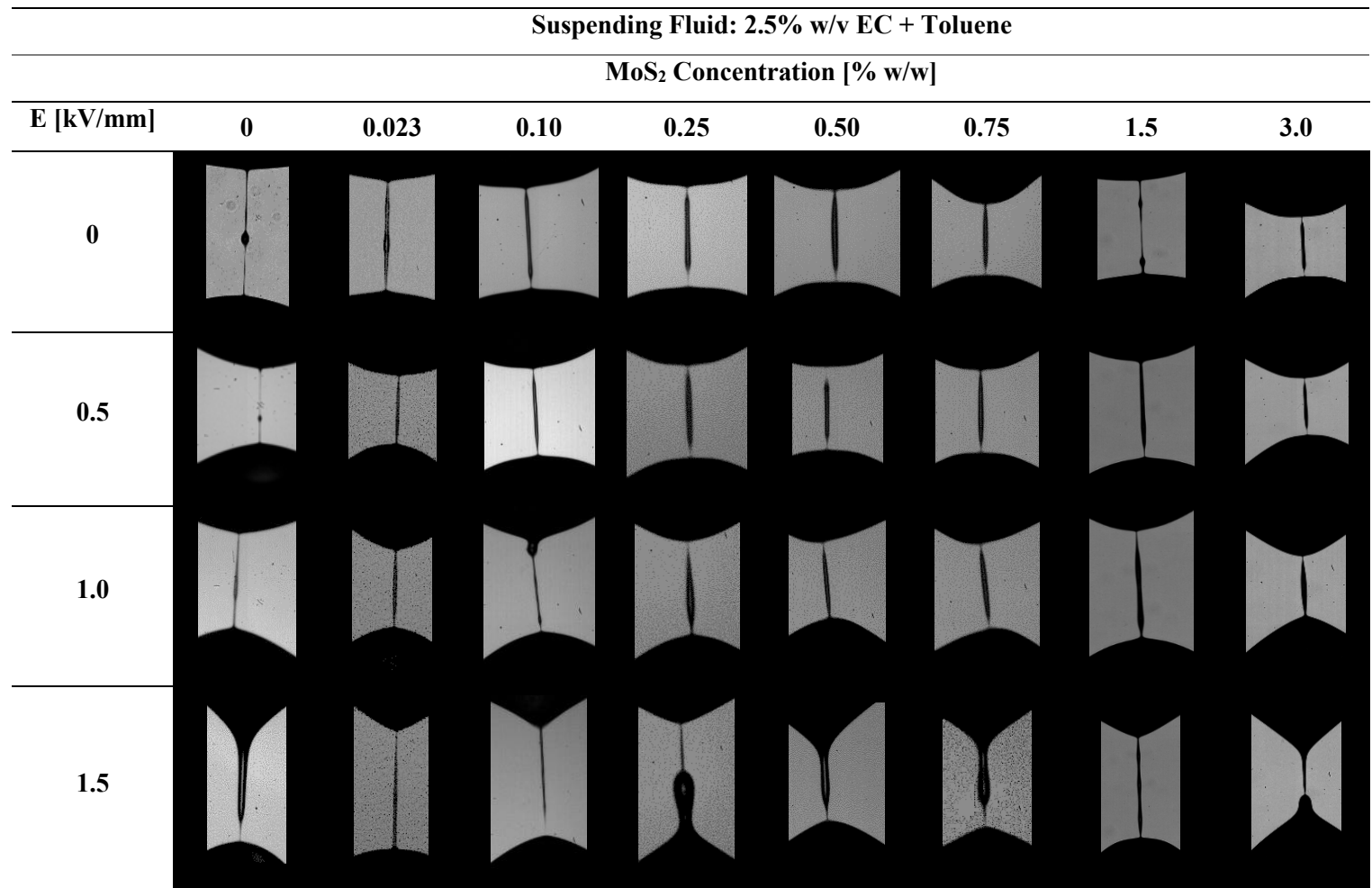

**Figure 10.** Last photograph for each MoS$_2$-ink immediately before the filament breakage.

When $E_0 = 1.5$ kV/mm, an incipient formation of a Taylor cone is observed and independent of particle concentration (**Figure 10**). According to la Mora [27] and Ramos and Castellanos [28], the shape of the Taylor cone depends on many factors that alter the geometry of the cone as can be seen schematically in **Figure 11**. Furthermore, these two works showed that the ratio between the dielectric constant of the two media (in this work they correspond only to air and the fluid confined in the two plates) can influence the angle measured at the tip of the cone and the pressure gradient at the interface between the two fluids can change the curvature of the cone walls.

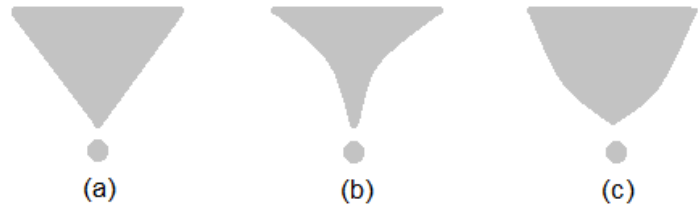

**Figure 11** – Schematic representation of the Taylor cone according to the pressure gradient: (a) $\Delta p = 0$, (b) $\Delta p < 0$ and (c) $\Delta p > 0$. Adapted from [27].

Given the experimental apparatus used in this work, we found that incipient Taylor cone formation occurs in the upper plate (positive polarity) for all suspensions studied here at 1.5kV/mm. However, it is observed that the geometry of the Taylor's cone depends on the $MoS_2$ concentration. For $MoS_2$ concentrations between 0.023% and 0.25%, and 1.5%, the Taylor cone formed is practically perfect (**Figure 11a**) and there is a prolongation of the tip of the cone (**Figure 11b**) for the remaining suspensions,. The presence or absence of the prolongation of the tip of the cone may be associated to the randomness of the particle and polymer concentration fluctuations in the filament when it is stretched. This latter observation may open a new use for the CaBEER [27], as quick tool assessing the formulations that can work well for electrohydrodynamic applications.

## 4. Conclusions

The application of an electric field to the polymeric solution promotes the appearance of vortices inside the fluid. This apparent increase in viscosity is due to the recirculation of fluid promoted by the vortices which is very similar to the phenomenon of secondary flows that appears at high shear rates in the absence of electric field, where the laminar flow condition is violated. When $MoS_2$ particles are added to the fluid, they migrate to the positive electrode, promoting the negative electrorheological effect since there is a phase separation of the particulate matter and the carrier fluid.

The negative electrorheological effect is more pronounced when the particle concentration is greater than 0.10%.

The dielectric properties of the MoS$_2$-inks show the impossibility of a positive electrorheological effect, i.e. an increase in viscosity due to the formation of columnar structures, since these fluids have a low polarization rate and a low difference of the dielectric constants in the range 100 to 10$^5$ Hz.

Regarding the extensional tests, it would be expected that the migration of MoS$_2$ particles to the positive electrode dragging with it the polymer presents in the fluid under the application of electric field. These two phenomena lead to significant changes in the rheological properties of the fluid, especially in the extensional relaxation time of the fluid. This relaxation time tends toward zero as the electric field strength increases. When the electric field strength is 1.5 kV/mm, the formation of Taylor cone is observed and independent of MoS$_2$ concentration.

**Acknowledgements**

Authors acknowledge FEDER, FCT/MCTES (PIDDAC) and FCT for funding support under grants UI/BD/150886/2021, NORTE-01-0145-FEDER-000054, LA/P/0045/2020, UIDB/00532/2020 and UIDP/00532/2020. Dr. Francisco Galindo-Rosales also acknowledges the funds from the program Stimulus of Scientific Employment, Individual Support 2020.03203.CEECIND. The authors also thank Prof. Cândido Duarte for helping in building up the setup to measure the dielectric properties of fluids.

**CRediT authorship contribution statement**

**Pedro C. Rijo:** Conceptualization, Methodology, Investigation, Formal analysis, Data curation, Writing – original draft.

**Francisco J. Galindo-Rosales:** Conceptualization, Methodology, Formal analysis, Resources, Supervision, Writing – review & editing, Funding acquisition.

# References

[1] D. Gupta, V. Chauhan, and R. Kumar, "A comprehensive review on synthesis and applications of molybdenum disulfide (MoS2) material: Past and recent developments," *Inorganic Chemistry Communications,* vol. 121, p. 108200, 2020/11/01/ 2020, doi: 10.1016/j.inoche.2020.108200.

[2] V. P. Kumar and D. K. Panda, "Review—Next Generation 2D Material Molybdenum Disulfide (MoS2): Properties, Applications and Challenges," *ECS Journal of Solid State Science and Technology,* vol. 11, no. 3, p. 033012, 2022/03/25 2022, doi: 10.1149/2162-8777/ac5a6f.

[3] C. Cong *et al.*, "Electrohydrodynamic printing for demanding devices: A review of processing and applications," vol. 11, no. 1, pp. 3305-3334, 2022, doi: 10.1515/ntrev-2022-0498.

[4] A. Fakhari, C. Fernandes, and F. J. Galindo-Rosales, "Mapping the Volume Transfer of Graphene-Based Inks with the Gravure Printing Process: Influence of Rheology and Printing Parameters," *Materials*, vol. 15, no. 7, doi: 10.3390/ma15072580.

[5] Z. Esa, M. Abid, J. H. Zaini, B. Aissa, and M. M. Nauman, "Advancements and applications of electrohydrodynamic printing in modern microelectronic devices: a comprehensive review," *Applied Physics A,* vol. 128, no. 9, p. 780, 2022/08/17 2022, doi: 10.1007/s00339-022-05796-3.

[6] F. Marra, S. Minutillo, A. Tamburrano, and M. S. Sarto, "Production and characterization of Graphene Nanoplatelet-based ink for smart textile strain sensors via screen printing technique," *Materials & Design,* vol. 198, p. 109306, 2021/01/15/ 2021, doi: 10.1016/j.matdes.2020.109306.

[7] C. Kamal, S. Gravelle, and L. Botto, "Hydrodynamic slip can align thin nanoplatelets in shear flow," *Nature Communications,* vol. 11, no. 1, 2020, doi: 10.1038/s41467-020-15939-w.

[8] R. Ivanova *et al.*, "Composition dependence in surface properties of poly(lactic acid)/graphene/carbon nanotube composites," *Materials Chemistry and Physics,* vol. 249, p. 122702, 2020/07/15/ 2020, doi: 10.1016/j.matchemphys.2020.122702.

[9] H. Wang *et al.*, "Improved preparation of MoS2/graphene composites and their inks for supercapacitors applications," *Materials Science and Engineering: B,* vol. 262, p. 114700, 2020/12/01/ 2020, doi: 10.1016/j.mseb.2020.114700.

[10] P. C. Rijo, J. M. O. Cremonezzi, R. J. E. Andrade, and F. J. Galindo-Rosales, "Correlation between the rheology of electronic inks and the droplet size generated from a capillary nozzle in dripping regime," *Physics of Fluids,* vol. 35, no. 9, p. 093116, 2023, doi: 10.1063/5.0166228.

[11] S. Lee, Y. K. Kim, J.-Y. Hong, and J. Jang, "Electro-response of MoS2 Nanosheets-Based Smart Fluid with Tailorable Electrical Conductivity," *ACS Applied Materials & Interfaces,* vol. 8, no. 36, 2016, doi: 10.1021/acsami.6b07887.

[12] M. Mrlík, M. Ilčíková, T. Plachý, R. Moučka, V. Pavlínek, and J. Mosnáček, "Tunable electrorheological performance of silicone oil suspensions based on controllably reduced graphene oxide by surface initiated atom transfer radical polymerization of poly(glycidyl methacrylate)," *Journal of Industrial and Engineering Chemistry,* vol. 57, pp. 104-112, 2018/01/25/ 2018, doi: 10.1016/j.jiec.2017.08.013.

[13] J. Yin, Y. Shui, R. Chang, and X. Zhao, "Graphene-supported carbonaceous dielectric sheets and their electrorheology," *Carbon,* vol. 50, no. 14, pp. 5247-5255, 2012/11/01/ 2012, doi: 10.1016/j.carbon.2012.06.062.

[14] P. C. Rijo and F. J. Galindo-Rosales, "The building blocks behind the electrohydrodynamics of non-polar 2D-inks," *Applied Materials Today,* vol. 36, p. 102042, 2024/02/01/ 2024, doi: 10.1016/j.apmt.2023.102042.

[15] A. Nojiri and S. Okamoto, "Dielectric Properties of Polymer–Solvent Systems," *Polymer Journal,* vol. 2, no. 6, pp. 689-697, 1971/11/01 1971, doi: 10.1295/polymj.2.689.

[16] F. Hussain, M. Hojjati, M. Okamoto, and R. E. Gorga, "Review article: Polymer-matrix Nanocomposites, Processing, Manufacturing, and Application: An Overview," *Journal of*

| [ ] | |
|---|---|
| | *Composite Materials,* vol. 40, no. 17, pp. 1511-1575, 2006/09/01 2006, doi: 10.1177/0021998306067321. |
| [17] | F. J. Galindo-Rosales, P. Moldenaers, and J. Vermant, "Assessment of the Dispersion Quality in Polymer Nanocomposites by Rheological Methods," *Macromolecular Materials and Engineering,* vol. 296, no. 3-4, pp. 331-340, 2011/03/28 2011, doi: 10.1002/mame.201000345. |
| [18] | Q. Wan, Y. Jin, P. Sun, and Y. Ding, "Rheological and tribological behaviour of lubricating oils containing platelet MoS2 nanoparticles," *Journal of Nanoparticle Research,* vol. 16, no. 5, p. 2386, 2014/04/08 2014, doi: 10.1007/s11051-014-2386-2. |
| [19] | J. K. G. Panitz, M. T. Dugger, D. E. Peebles, D. R. Tallant, and C. R. Hills, "Electrophoretic deposition of pure MoS2 dry film lubricant coatings," *Journal of Vacuum Science & Technology A,* vol. 11, no. 4, pp. 1441-1446, 1993/07/01 1993, doi: 10.1116/1.578570. |
| [20] | A. Barrero, A. M. Gañán-Calvo, J. Dávila, A. Palacios, and E. Gómez-González, "The role of the electrical conductivity and viscosity on the motions inside Taylor cones," *Journal of Electrostatics,* vol. 47, no. 1, pp. 13-26, 1999/06/01/ 1999, doi: 10.1016/S0304-3886(99)00021-2. |
| [21] | K. Mohammadi, M. R. Movahhedy, and S. Khodaygan, "Colloidal particle reaction and aggregation control in the Electrohydrodynamic 3D printing technology," *International Journal of Mechanical Sciences,* vol. 195, p. 106222, 2021/04/01/ 2021, doi: 10.1016/j.ijmecsci.2020.106222. |
| [22] | L. Campo-Deaño and C. Clasen, "The slow retraction method (SRM) for the determination of ultra-short relaxation times in capillary breakup extensional rheometry experiments," *Journal of Non-Newtonian Fluid Mechanics,* vol. 165, no. 23-24, pp. 1688-1699, 2010, doi: 10.1016/j.jnnfm.2010.09.007. |
| [23] | J. H. Ock, J. S. Hong, and K. H. Ahn, "Acceleration of instability during the capillary thinning process due to the addition of particles to a poly(ethylene oxide) solution," *Journal of Non-Newtonian Fluid Mechanics,* vol. 258, pp. 58-68, 2018/08/01/ 2018, doi: 10.1016/j.jnnfm.2018.04.009. |
| [24] | W. Mathues, C. McIlroy, O. G. Harlen, and C. Clasen, "Capillary breakup of suspensions near pinch-off," *Physics of Fluids,* vol. 27, no. 9, 2015, doi: 10.1063/1.4930011. |
| [25] | G. H. McKinley and A. Tripathi, "How to extract the Newtonian viscosity from capillary breakup measurements in a filament rheometer," *Journal of Rheology,* vol. 44, pp. 653-670, 2000, doi: 10.1122/1.551105. |
| [26] | C. Clasen, "Capillary breakup extensional rheometry of semi-dilute polymer solutions," *Korea-Aust Rheol J,* vol. 22, pp. 331-338, 12/01 2010. |
| [27] | J. Fernández de la Mora, "The Fluid Dynamics of Taylor Cones," *Annual Review of Fluid Mechanics,* vol. 39, no. 1, pp. 217-243, 2007/01/01 2006, doi: 10.1146/annurev.fluid.39.050905.110159. |
| [28] | A. Ramos and A. Castellanos, "Conical points in liquid-liquid interfaces subjected to electric fields," *Physics Letters A,* vol. 184, no. 3, pp. 268-272, 1994/01/10/ 1994, doi: 10.1016/0375-9601(94)90387-5. |

Supporting Information

# Rheology of MoS$_2$ inks: Influence of MoS$_2$ Concentration in the Presence of Electric Field

*Pedro C. Rijo and Francisco J. Galindo-Rosales*

## I. Dielectric Constant and Dielectric Loss of MoS$_2$-inks

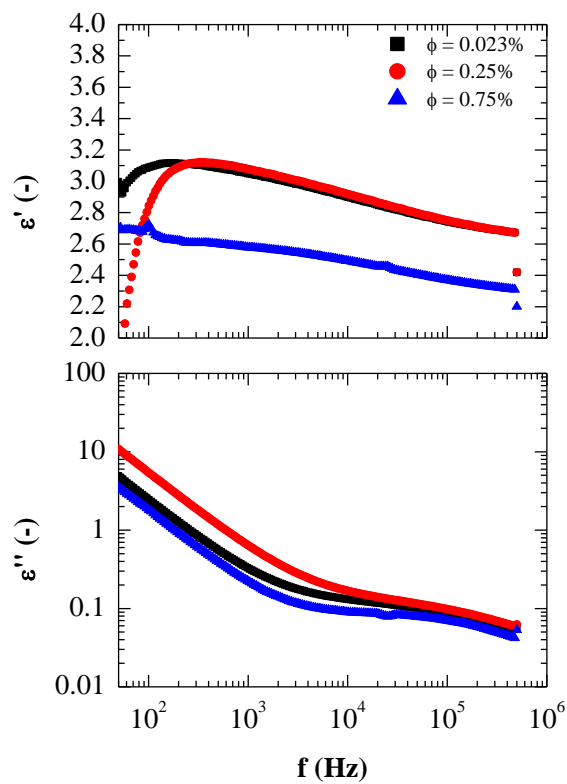

**Figure S 1**. Dielectric constant ($\varepsilon'$) and dielectric loss factor for five different MoS$_2$-inks ($\phi$ = 0.023%; 0.25% and 0.75%).

## II. Rheology and Electrorheology under shear flow condition

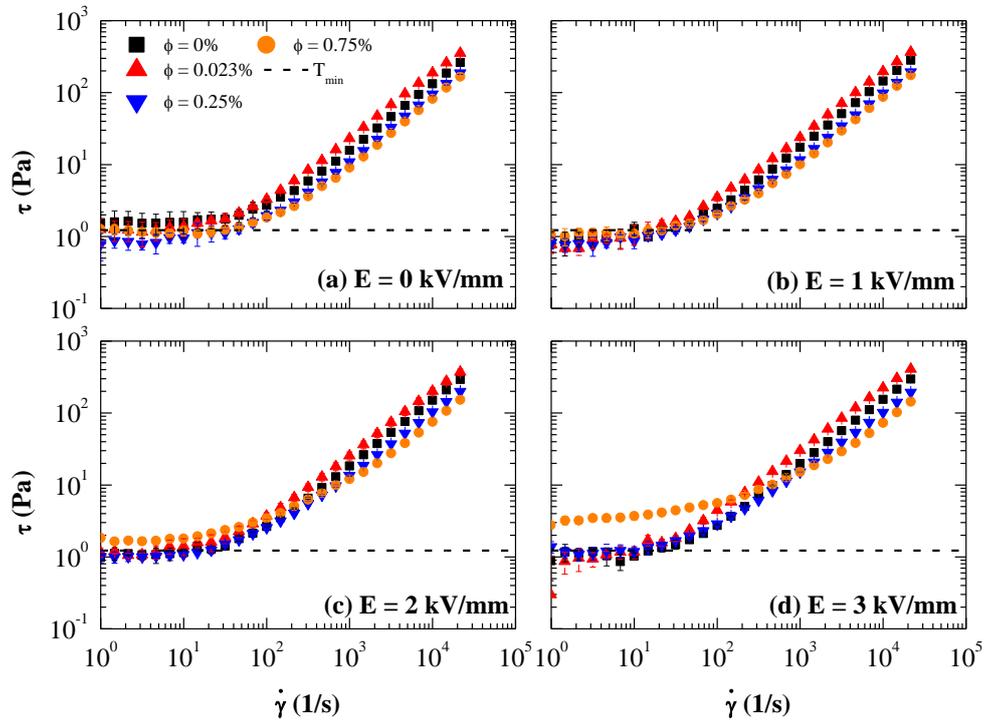

**Figure S 2.** Influence of MoS$_2$ concentration (ϕ = 0ppm, 0.023%; 0.25% and 0.75%). on the shear stress curves for: **(a)** E = 0 kV/mm, **(b)** E = 1 kV/mm, **(c)** E = 2 kV/mm, and **(d)** E = 3 kV/mm.

**Video S1** shows the particle migration of MoS$_2$ for a particle concentration of 0.50% w/w. The MoS$_2$ paricles were dispersed in 2.5% w/v of ethyl cellulose+toluene under an electric field strength of 1.5 kV/mm. The distance between the two electrodes was 2 mm.

## III. Rheology and Electrorheology under uniaxial flow condition

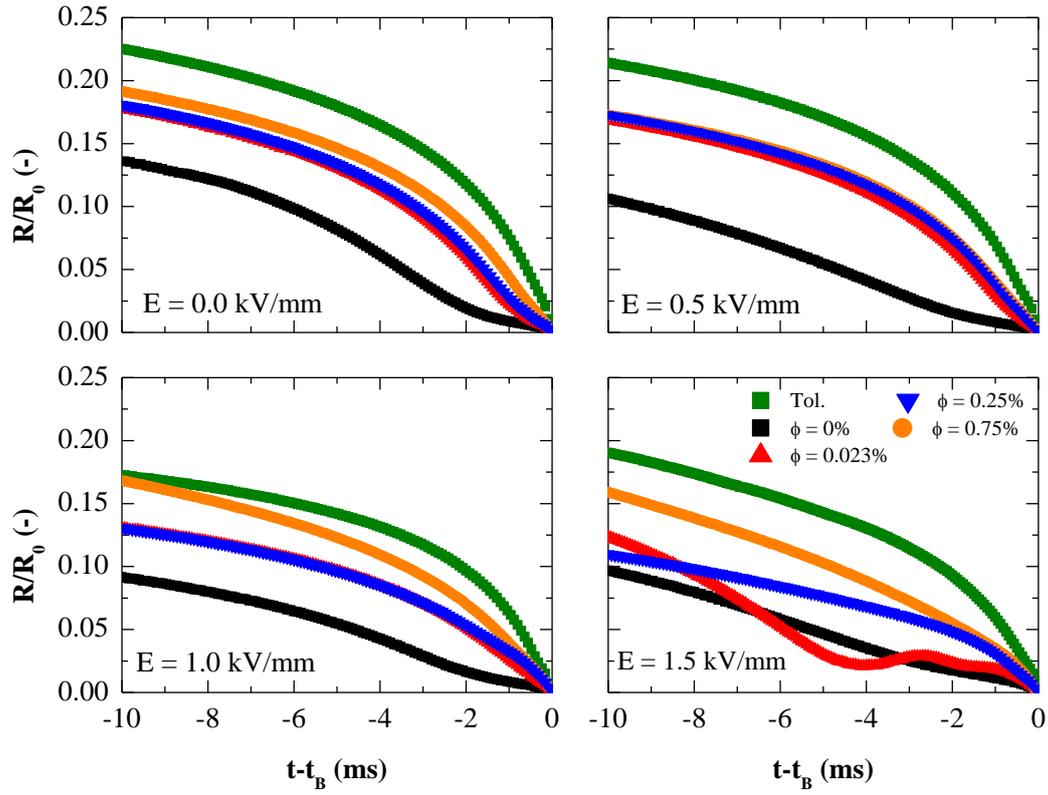

**Figure S 3.** The normalized radius decay profiles for five different MoS$_2$-inks ($\phi$ = 0%, 0.023%, 0.25% and 0.75% w/w) and toluene for different electric field strengths: **(a)** 0 kV/mm, **(b)** 0.5 kV/mm, **(c)** 1.0 kV/mm and **(d)** 1.5 kV/mm.

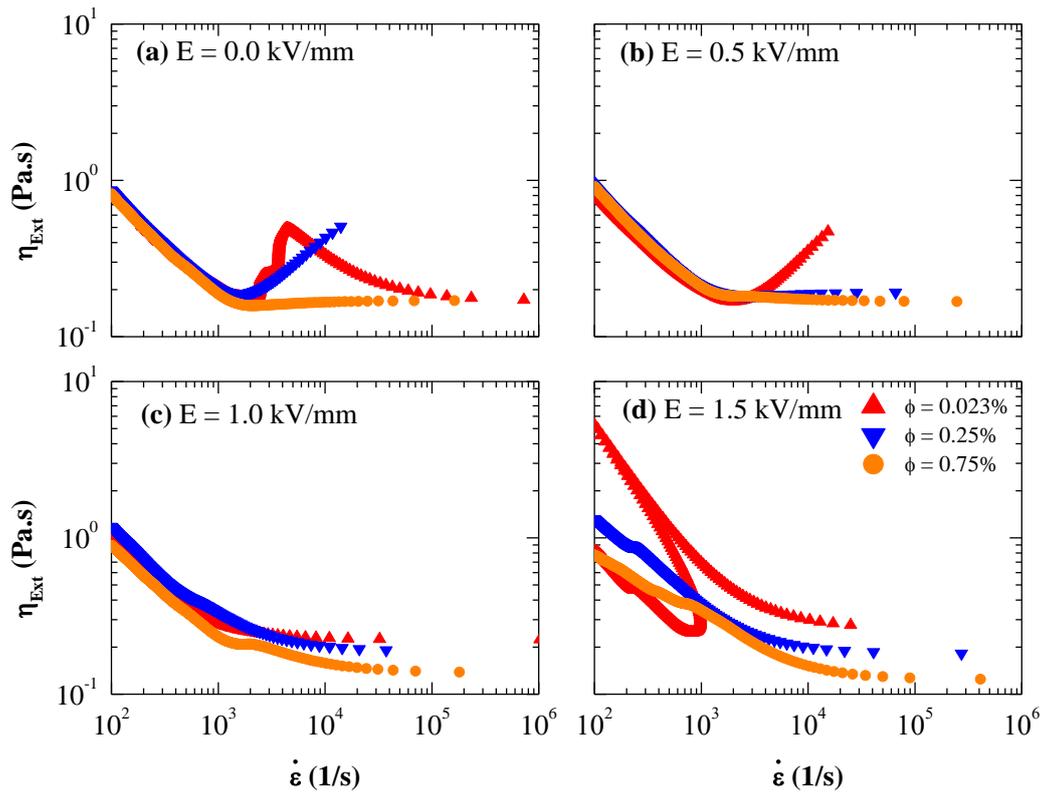

**Figure S 4.** The apparent extensional viscosity ($\eta_{Ext}$) as a function of extension rate ($\dot{\varepsilon}$) for three different MoS$_2$-inks ($\phi$ = 0.023%, 0.25% and 0.75% w/w) for different electric field strengths: **(a)** 0 kV/mm, **(b)** 0.5 kV/mm, **(c)** 1.0 kV/mm and **(d)** 1.5 kV/mm.

The photographic record of the thinning process for the carrier fluid (2.5% w/v EC+Toluene) and MoS$_2$-ink with a concentration of 0.023% w/w are available in the work done by Rijo and Galindo-Rosales [1]. **Figure S 5** to **Figure S 10** show the photographic record of the thinning process for MoS$_2$-inks with a concentration of 0.10% w/w, 0.25% w/w, 0.50% w/w, 0.75% w/w, 1.5% w/w and 3.0% w/w, respectively.

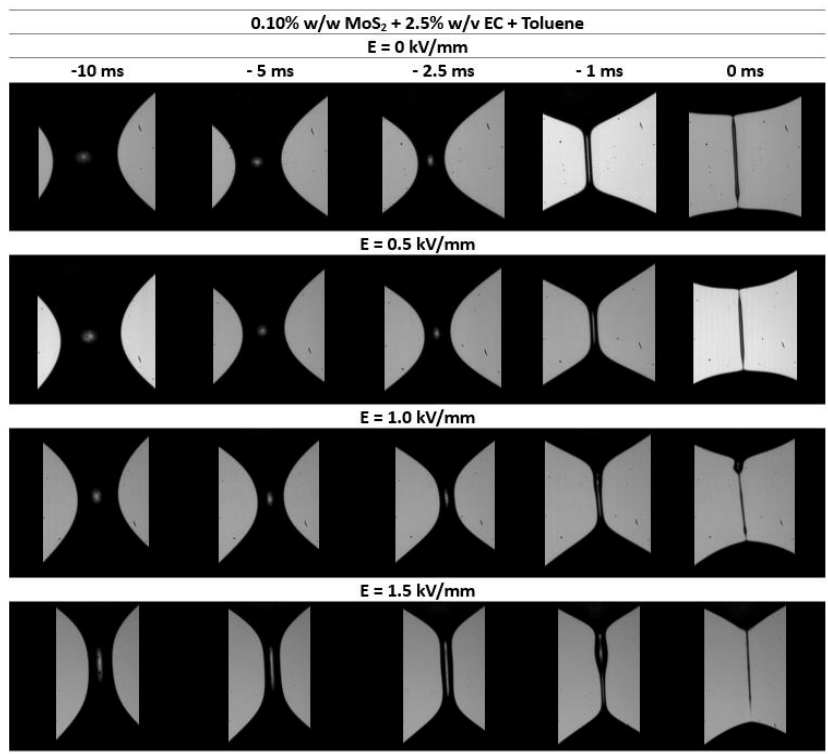

**Figure S 5.** Photographic record of the thinning process for MoS$_2$-ink has concentration of 0.10% w/w for several electric field strengths.

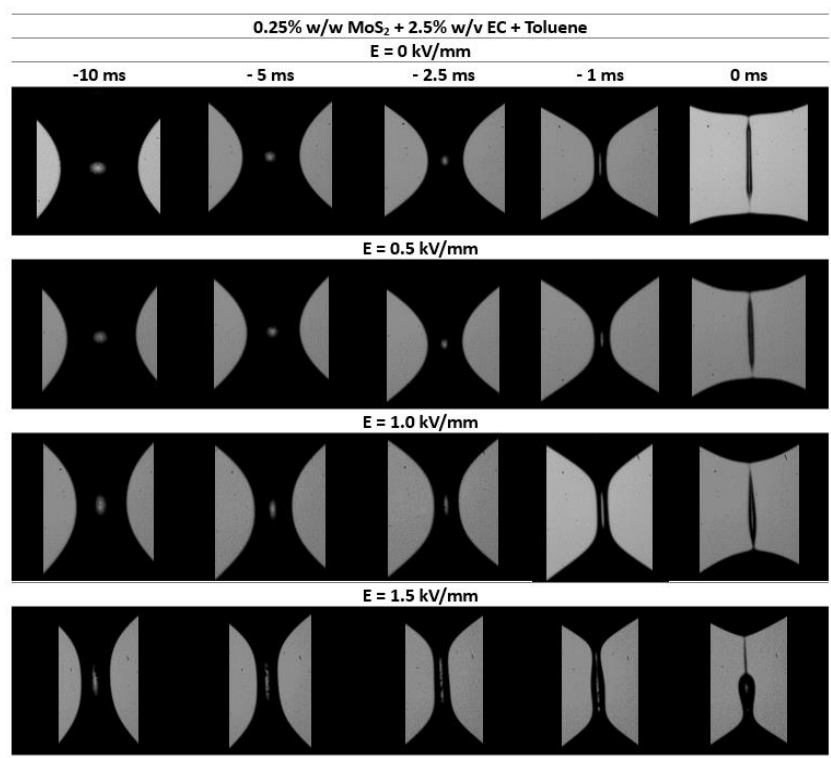

**Figure S 6.** Photographic record of the thinning process for MoS$_2$-ink has concentration of 0.25% w/w for several electric field strengths.

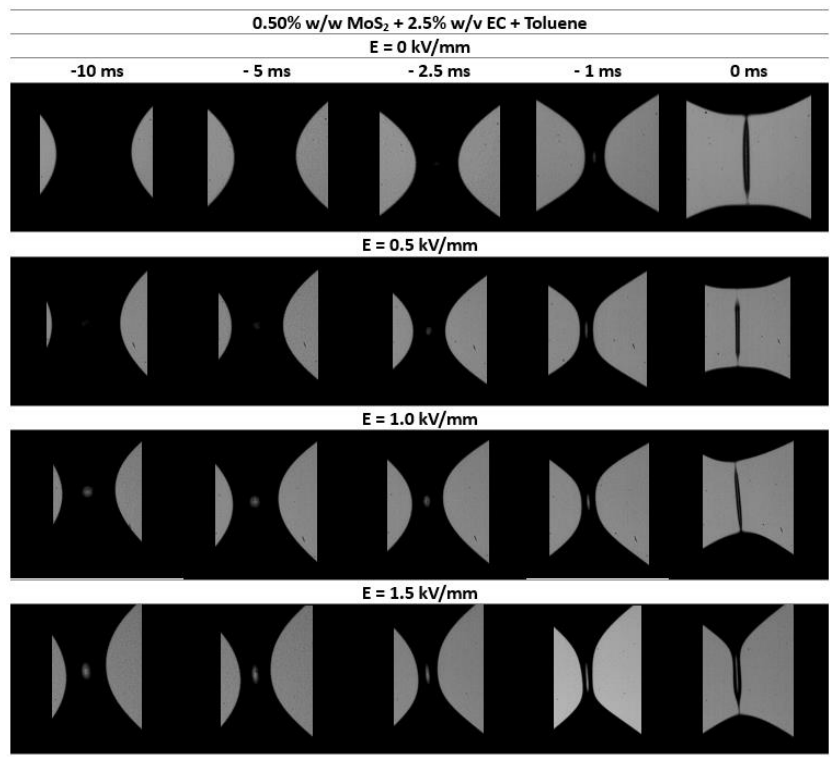

**Figure S 7.** Photographic record of the thinning process for MoS$_2$-ink has concentration of 0.50% w/w for several electric field strengths.

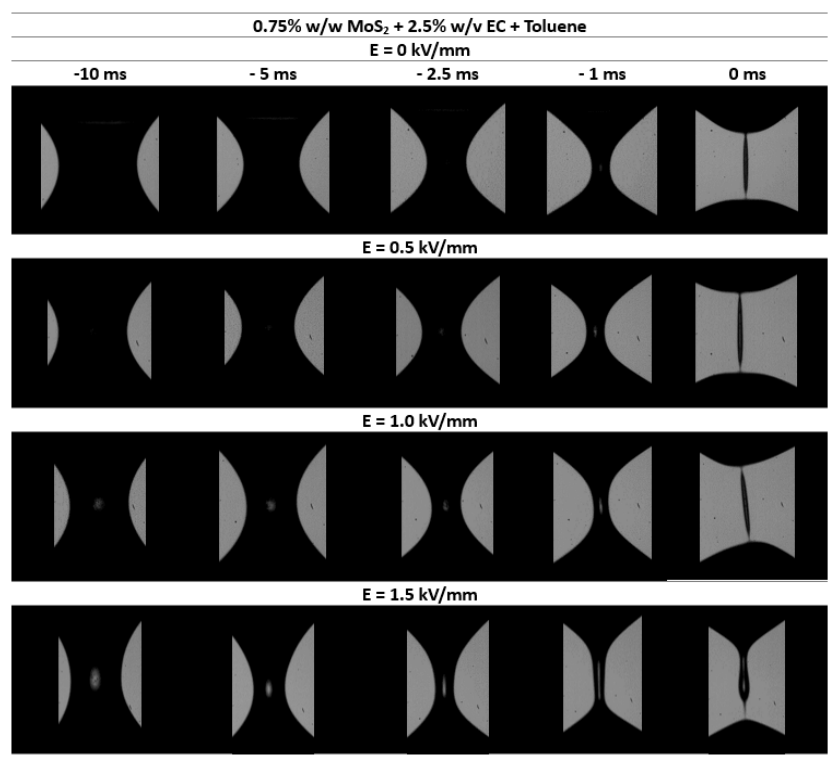

**Figure S 8.** Photographic record of the thinning process for MoS$_2$-ink has concentration of 0.75% w/w for several electric field strengths.

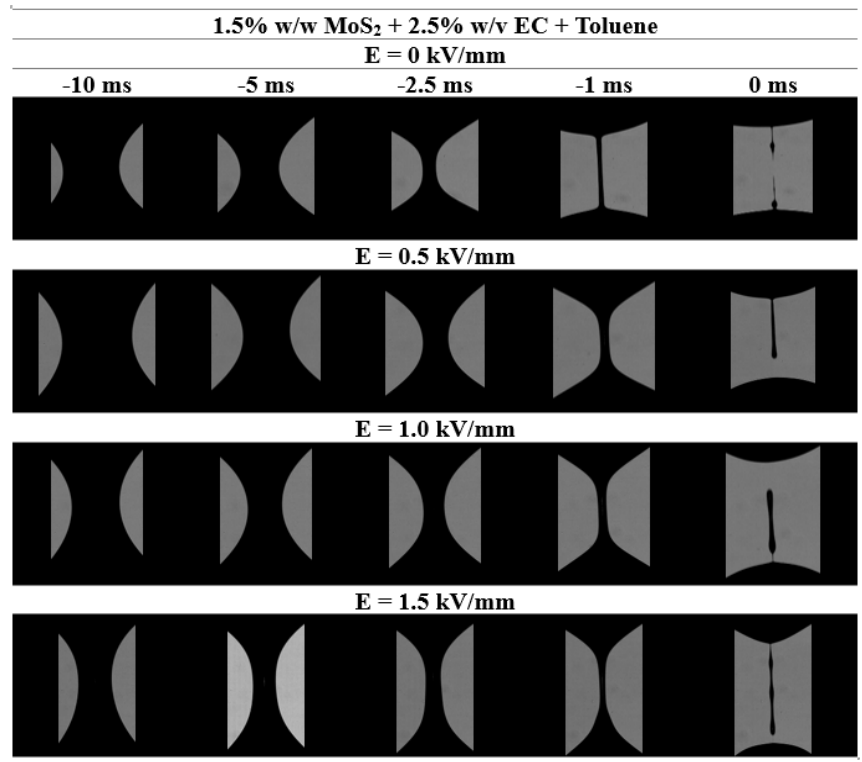

**Figure S 9** Photographic record of the thinning process for MoS$_2$-ink has concentration of 1.5% w/w for several electric field strengths.

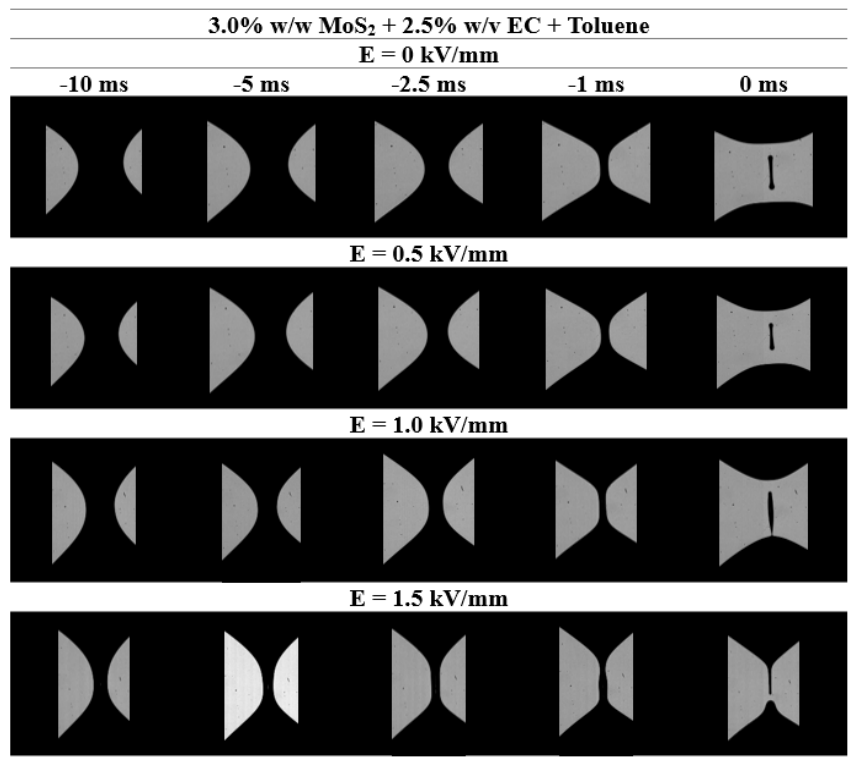

**Figure S 10.** Photographic record of the thinning process for $MoS_2$-ink has concentration of 3.0% w/w for several electric field strengths.

## References

[1] P. C. Rijo and F. J. Galindo-Rosales, "The building blocks behind the electrohydrodynamics of non-polar 2D-inks," Applied Materials Today, vol. 36, p. 102042, 2024/02/01/ 2024, doi: 10.1016/j.apmt.2023.102042.